%% file: main.tex
\documentclass[runningheads]{llncs}

\usepackage[utf8]{inputenc}
\usepackage[T1]{fontenc}

\usepackage{newtxtext}
\usepackage[varvw]{newtxmath}

\usepackage{multirow}
\usepackage{booktabs}
\usepackage{tikz}
\usepackage{tabularx}
\usepackage{array}
\usepackage{graphicx}

\usetikzlibrary{arrows.meta, positioning, shapes.geometric}

\begin{document}

\begin{center}
\small\itshape \textbf{This is a preprint version. The final version will appear in the proceedings of PROFES~2026.}
\end{center}

\title{Rethinking Data Quality for AI-Driven Systems: Evidence from Practitioner Interviews}

\titlerunning{Rethinking Data Quality for AI-Driven Systems}

\author{Hariharan Gopinath\inst{1} \and
Jan Bosch\inst{1,2} \and
Helena Holmström Olsson\inst{3}}

\authorrunning{H. Gopinath et al.}

\institute{
Chalmers University of Technology, Department of Computer Science and Engineering, Gothenburg, Sweden \\
\email{hargop@chalmers.se, jan.bosch@chalmers.se}
\and
Eindhoven University of Technology, Department of Mathematics and Computer Science, Eindhoven, Netherlands \\
\and
Malm\"o University, Department of Computer Science and Media Technology, Malm\"o, Sweden \\
\email{helena.holmstrom.olsson@mau.se}
}
\maketitle
\begin{abstract}

Data quality research has usually treated data as an input that is stored, processed, and validated. In AI-driven software-intensive systems, data also shapes model behavior, evaluation, and lawful use. Empirical evidence remains limited on how practitioners define, assess, and manage quality under these conditions. We interviewed 16 practitioners from nine organizations and analyzed the transcripts using reflexive thematic analysis and developed six themes from participants’ accounts. In AI systems, traceability shifted from modular debugging to attributing model behavior, while using models as quality assessors introduced circularity. Agent context and memory became data objects, and synthetic and pseudo-labeled data made authenticity a quality concern. In foundation-model development, lawfulness became a gate for training data, while representativeness was judged through coverage of situations in which the system must behave safely. Prior ML research examines many of these problems separately. Our study provides a practitioner-grounded account of how they are encountered together as an engineering and organizational concern. We also interpret five recurring conditions as helping explain how the themes relate to reduced trust in data and AI outcomes. We synthesize these findings through lifecycle assurance: a conceptual framing focused on producing evidence that data can support a specific AI claim when its influence may be embedded in model behavior, model-based judgments, or agent actions.


\end{abstract}
\section{Introduction}
\label{sec:intro}
Data quality has long been understood as more than the absence of errors. Foundational work defines it as fitness for use: the same dataset may be adequate for one consumer, task, or decision and inadequate for another \cite{wang1996}. Traditional data management operationalizes this through dimensions such as accuracy, completeness, consistency, timeliness, accessibility, and provenance \cite{Batini}. These dimensions remain important, but AI-driven software-intensive systems change how data quality affects system behavior. Data is no longer only stored, queried, transformed, and validated, but it is also used to train models, evaluate behavior, populate runtime context, and constrain whether a system can be lawfully released or operated.

Production ML research already documents hidden data dependencies, feedback loops, and blurred system boundaries \cite{Hidden_Technical_Debt}, as well as data work that continues from collection through monitoring \cite{SE_for_ML_lifecycle,Data_Management_Challenges_Production_ML}. Weak data practices can also propagate into unreliable models and harmful outcomes \cite{Sambasivan2021EveryoneWT}. These studies usually examine provenance, synthetic data, documentation, or governance separately. We know less about how practitioners face these concerns together while training, evaluating, deploying, and operating AI-driven systems. We therefore interviewed 16 practitioners from nine organizations working in automotive, telecom, data annotation, industrial automation, decentralized AI, and foundation-model development. The interviews addressed three research questions (RQs):

\begin{description}
\item[RQ1.] How do practitioners define and assess data quality in AI-driven systems, and how does this differ from traditional data management?
\item[RQ2.] What data quality challenges do practitioners encounter across the AI development lifecycle?
\item[RQ3.] How do organizational structure, technical factors, and emerging AI artifacts shape practitioners' trust in data for AI-driven systems?
\end{description}

 In this paper, \emph{data trust} means justified confidence that data are suitable for a specific AI use and that their origin, transformations, ownership, permitted use, and influence on system behavior can be examined and accounted for. It is one part of AI system trustworthiness, alongside model design, human oversight, interfaces, and organizational controls. 
 
Our analysis produced six practitioner-grounded themes showing how AI reshapes data-quality practice (Sect.~\ref{sec:Findings}). We interpret five foundational explanatory mechanisms that help explain how these changes weaken trust in data and AI outcomes (Sect.~\ref{sec:foundational_causes}). We synthesize these findings as lifecycle assurance: producing artifact-specific evidence that data is fit for claims made during training, validation, certification, deployment, and agent action (Sect.~\ref{sec:5.3}).

\section{Background and Related Work}
\label{sec:background}
\subsection{Background}

Data quality has traditionally been understood as a purpose-dependent concept. Wang and Strong define data quality as “fitness for use” meaning that the same data may be suitable for one consumer, task, or decision but unsuitable for another~\cite{wang1996}. This view is broader than accuracy alone. It includes whether data is complete, timely, interpretable, accessible, and relevant to the context in which it is used. Later work on data-quality assessment similarly distinguishes measurable properties of data from user- and context-dependent judgments of quality~\cite{pipino2002}. For this study, this literature provides the baseline, as data quality is not a universal property of a dataset but a judgment about whether data can support a particular use.

Traditional data-quality research also draws on provenance and governance. Provenance records where data originated and how it was transformed, supporting interpretation and reproducibility~\cite{simmhan2005}. Data governance defines decision rights and accountability for quality, access, metadata, and lifecycle management~\cite{khatri2010}. For AI-driven systems, these perspectives direct attention beyond record contents to an organization's ability to explain where data came from, how it changed, and who is responsible for its use. Provenance and ownership provide concepts for examining how practitioners justify trust in data. Our study investigates how this concern extends from data records to model behavior.

Machine learning complicates this account because data enters system behavior instead of remaining stored, queried, or reported. Research on technical debt shows that ML systems create hidden data dependencies, feedback loops, unstable boundaries, and downstream consumers that may be difficult to anticipate~\cite{Hidden_Technical_Debt}. Production ML research also places data work throughout collection, cleaning, labeling, feature preparation, training, testing, deployment, and monitoring~\cite{Data_Management_Challenges_Production_ML}. Data quality is consequently a lifecycle concern: decisions made upstream can later affect model behavior, evaluation results, and operational reliability.

\subsection{Related Work}

Empirical studies of AI development show that data problems propagate. Sambasivan et al. describe "data cascades" in high-stakes AI, where undervalued data work produces downstream model failures~\cite{Sambasivan2021EveryoneWT}. Their study examines poor data practices and their consequences. We ask how practitioners define, assess, and trust data quality across end-to-end, foundation-model, and agentic systems.
Data-centric AI argues that improving AI systems requires systematic attention to data, not only model architecture~\cite{data_centric}. Datasheets for Datasets, likewise, makes composition, collection, and intended use explicit~\cite{Datasheets_for_Datasets}. We extend this evidence-based view beyond source datasets to pseudo-labels, synthetic data, model judgments, agent context, and legal provenance. Qian et al.'s interviews with ten dataset practitioners in one organization found that quality was a priority but lacked a shared definition and was often assessed through ad hoc inspection~\cite{Understanding_the_Dataset_Practitioners}. 

Further, LLM-as-a-judge research shows that model-based evaluators carry their own biases \cite{zheng2023}, recursive training degrades models trained on their own output \cite{shumailov2024}, and provenance audits find licensing hard to establish at training-data scale \cite{dataprovenance_Attribution_AI}. Prior work documents these problems individually, but less is known about how practitioners encounter and connect them as a data-quality and trust problem across the AI lifecycle. We address this gap through an integrated practitioner account and synthesize the findings as lifecycle assurance: producing evidence that data artifacts are fit for the specific AI claims they support.

\section{Research Method}
\label{sec:method}

We conducted an exploratory interview study with 16 practitioners from nine organizations involved in AI development, deployment, or data provision. We used purposive sampling to recruit information-rich participants with direct experience of AI development, deployment, or data provision, and complemented this with snowball sampling through participant and professional network referrals \cite{palinkas2015purposeful}. The sample was designed to capture different organizational relationships with AI, including organizations that train and deploy their own models, organizations that provide data or annotation services, and organizations applying foundation models in enterprise settings. Participants are anonymized as I01-I16 and their nine organizations are grouped into four domain areas in Table \ref{tab:participants}.

The sample was intentionally weighted toward safety-critical perception-and-control AI development, where autonomous driving makes data quality especially demanding. To broaden the empirical basis, we added participants working on foundation models, decentralized and federated AI, large-scale data services, industrial automation, and consultancy. \input{tables/participants}

\subsection{Data Collection}

The interviews were conducted between April 2026 and June 2026 in English, either through Microsoft Teams or on site at participating organizations. With participants' consent, all interviews were audio-recorded. Before each interview, participants were informed that their names, organizations, and personal information would be anonymized in research outputs. Audio recordings and full transcripts other than supplementary material would not be shared or used for purposes beyond this study. Each interview lasted between 45 and 60 minutes and was semi-structured. The interview guide was designed around the three research questions discussed in Section~\ref{sec:intro}. Semi-structured interviews were appropriate because they provided a common structure across participants while allowing follow-up questions and concrete examples from participants’ own work~\cite{semi_interview}.

Interview recordings were transcribed using a locally installed version of WhisperX~\cite{whisperx} on the first author's computer. The recordings were not uploaded to or processed by external AI services. WhisperX generated the initial transcripts, aligned words with the audio, performed speaker diarization, and exported speaker-labeled transcripts with timestamps. The first author then reviewed each transcript against the original audio and manually corrected transcription errors, especially domain-specific terms and phrases where mistakes changed the meaning. During transcription and analysis, AI was used only for transcription and speaker separation. Coding, theme construction, cross-theme interpretation, and analytical writing were performed by the authors.

\subsection{Data Analysis}

We analyzed the transcripts using reflexive thematic analysis (RTA), following Braun and Clarke's six-phase approach \cite{TA}. RTA suited this exploratory study because it allowed us to examine how practitioners define, assess, and manage data quality in AI-driven systems without imposing predefined data-quality categories. First, in familiarization, the first author read the transcripts repeatedly and wrote analytical notes on recurring concerns, contradictions, and concrete examples. Second, generating initial codes, relevant transcript segments were coded inductively in Taguette, capturing how practitioners described and assessed data quality across the AI lifecycle. For instance, extracts such as "no reasoning trace" and "pixel to torque" were coded as behavior-level attribution collapse. Third, searching for themes, related codes were consolidated and grouped into candidate themes by comparing patterns across participants, organizations, and lifecycle stages. The code above, with related ones, formed the candidate theme \textit{"Behavior-Level Traceability"}. Fourth, reviewing themes, the candidate themes were checked against the coded extracts and the full dataset, and codes were merged, separated, or repositioned when they described different aspects of the same issue. Fifth, defining and naming themes: each theme was named to capture its analytical meaning rather than only its topic, and the sixth phase, producing the report, presents them in Section~\ref{sec:Findings}.

After constructing the six themes, we compared patterns across them in a second interpretive step. This analysis identified five foundational explanatory mechanisms that helped us interpret why several themes recurred and how they related to practitioners' accounts of diminished trust. For instance, the attribution problems in Theme 1, such as "pixel to torque" and the absence of a reasoning trace, and the coverage gaps that became driving behavior in Theme 6 share an underlying condition that data does not remain an inspectable input but is absorbed into model behavior. We consolidated this condition as F3. This cross-theme analysis resulted in the five foundational explanatory mechanisms (F1-F5) presented in Section~\ref{sec:foundational_causes}. These are not additional themes but second-order explanations that connect the six themes and support the model in Figure~\ref{fig:causes}. The analysis followed the chain:
\begin{center}
 \emph{quote} $\rightarrow$ \emph{initial code} $\rightarrow$ \emph{consolidated code} $\rightarrow$ \emph{theme} $\rightarrow$ \emph{explanatory mechanism}
\end{center}

We created a supplementary evidence table to keep the path from the empirical material to our interpretations visible. It links quote and participant IDs to initial codes, consolidated codes, themes, and explanatory mechanisms. The six themes draw on 45 quote-level extracts, including one used for two themes. The five mechanisms draw on 15 additional extracts and seven also used for themes. These counts document analytical grounding, not frequency. The full anonymized table is available on Zenodo\footnote{\url{https://doi.org/10.5281/zenodo.22901446}}. Following reflexive thematic analysis, we treated coding as interpretive and did not calculate inter-coder agreement.


\section{Findings}
\label{sec:Findings}
The thematic analysis produced six themes (T) through inductive coding. Rather than mapping directly onto individual research questions, the themes collectively inform all three, and we draw them together to answer each research question in Section~\ref{sec:discussion}.

\subsection*{T1: Behavior-Level Traceability}
\label{subsec:theme1}
\textit{The shift: Participants traced model behavior back through the data and pipeline that produced it.}

In traditional systems, traceability connects requirements, code, transformations, tests, and defects~\cite{clelandhuang2014}, and lineage tracks records through pipelines~\cite{simmhan2005}. In end-to-end AI systems, practitioners also trace behaviors and autonomous actions back through model versions, training data distributions, source data, and labels. One participant (I12) contrasted modular pipelines, where perception, planning, and control had clear interfaces, with end-to-end models that are far harder to debug. A model may learn correlations rather than causal relations, associating a red light with forward movement or acquiring a driving bias from an imbalanced training mix, consistent with causal misidentification in imitation learning~\cite{dehaan2019causalconfusionimitationlearning} and findings that hidden dataset biases can be absorbed into end-to-end driving behavior~\cite{zimmerlin2024hiddenbiasesendtoenddriving}. Because such models offer no reasoning trace and are, in the participant's words, "just a bunch of tensors" (I12), diagnosis depends on experimentation and A/B testing.

Others described the same collapse. One called it a "pixel to torque" problem (I01): when a vehicle acts incorrectly, it is hard to know whether pixels, training data, or model internals caused it. As end-to-end boundaries expand, "the notion of data gets diluted" (I06), and even ground truth becomes behavioral, since teams must judge whether recorded human driving was good behavior to learn (I13). This makes lineage central: practitioners need to trace from "decision to model to data" (I10), because "any data without lineage is not trustable" (I07). Failures are not untraceable, but attribution becomes costly and indirect. Part of the difficulty is organizational, as teams own individual sources, models, or checks, while no role owns the chain connecting them, so a defect can cross several correct-looking stages before it surfaces (I03, I04). While ML research treats causal confusion and dataset bias as technical phenomena, our findings show practitioners experiencing them as an organizational traceability problem, where the unit of traceability expands from data records to model behavior.

\subsection*{T2: Models as Data-Quality Judges}
\textit{The shift: Participants used models to assess training data when rules and human review did not scale, introducing circularity.}

Assessment still relies on rule-based checks, sampling, and human review~\cite{pipino2002,wang1996}. In annotation-heavy automotive pipelines, manual inspection persists because some defects escape automated checks. For example, if a road-lane marking is placed incorrectly in lidar point-cloud data, the error "cannot be checked by the sanity check script," yet there is "no way to check every one of them" (I05).
Because manual review does not scale, the same engineer turned the model into the assessor. After training an initial model, the team runs inference over all data and treats an unusually high loss as a signal that "actually the annotation of this specific sequence is wrong," reworking those sequences before the next cycle (I05). This surfaces mislabeled data at a scale no human review could reach, but it is circular. The judging model was trained on annotations of the same kind and inherits their blind spots, and a defect it was never taught to recognize passes silently back into training. Another participant named the pattern "model as an assessor, model as a judge" (I01).

The circularity deepens when the assessor's own data is model-generated, creating what one participant called "an infinite loop of models training on their own data" that severs external ground truth (I15). Practitioners therefore insisted that "we cannot automate everything" and that human "check and balance" remains essential (I11). The question "Is my data good?" becomes "Is my checker's data good?" The ML literature documents related risks such as self-preference and biases in model-based evaluation~\cite{zheng2023}. Our contribution is the organizational view: assurance adopted out of necessity can create a recursive accountability gap that is absent when validation remains external to the model.

\subsection*{T3: Agent Context and Memory as a New Data Object}

\textit{The shift: Participants treated retrieved context and retained memory as part of the agent's operational data layer.}

Access control, retention, and governance are familiar enterprise problems~\cite{khatri2010}. With agentic systems, data is also retrieved into context, retained in memory, reused, and acted upon, so context and memory become quality objects in their own right.
A senior AI lead (I08) framed the precondition: organizational data must be exposed "over an API in structured data" so "agents can consume it." Governance then becomes a real-time problem, since manual access approval "didn't work before, but it certainly doesn't work going forward" (I08), raising the question of who is accountable when agents act on behalf of people.

A data and AI architect (I10) made the failure mode concrete. As an agent writes new context on top of old, it produces "a new silo" that "was not factored in before in the whole production life cycle". That memory can be corrupted, as when a coding agent's memory became "infiltrated" and the agent responded strangely until the memory was cleared and the codebase reloaded (I10). The defect lives in accumulated operational context, not in any source dataset, and silently changes decisions. Another participant distrusted agentic outputs simply because "I don't know where it comes from" (I16).

We report this theme as emergent rather than saturated as only three participants discussed agent memory in depth, as agentic systems were still emerging in the participating organizations. One participant noted some data problems are unchanged in principle by AI (I08), and the researchers are beginning to govern agent memory technically~\cite{agentmemory2026,coala2024}. Practitioners report the organizational gap, that no role owns this layer, leaving provenance, freshness, and contamination unmanaged.

\subsection*{T4: Authenticity of Generated Data}

\textit{The shift: Whether data was observed or generated now affects whether it is fit for a particular use.}

Traditional frameworks asked whether derived data was accurate, representative, and fit for purpose~\cite{wang1996}. In AI-driven systems, models generate, annotate, and recycle data at scale, so data may be synthetic, pseudo-labeled, model-generated, or human-verified, and authenticity becomes a practical dimension: was the data observed from the world or produced by a model, and is that authentic enough for its lifecycle role?
A decentralized-AI participant illustrated this limitation with a classification example (I11). When synthetic data is generated from only a few real samples, “everything that is synthetic is around” those samples. If the real points occupy only one region of the space, the synthetic points also cluster there and fail to “cover the boundaries” between classes. The dataset therefore appears larger, but the decision boundary remains weakly observed. Related concerns appeared in foundation-model pseudo-annotation, where “the majority” of labels only needs to be good enough (I05), and in doubts about whether neural-rendered data captures real conditions (I06). For foundation-model training, AI-generated text was described as “not authentic data,” or at best “silver data,” creating “an infinite loop of models training on their own data” (I15), consistent with "model collapse"~\cite{shumailov2024}. 

The contrasting cases qualify the concern. One participant trusted synthetic depth more than measured outdoor depth because real estimates were unreliable, while synthetic scenes were “generated perfectly” (I09). Another used artificial scenarios to cover rare cases that would be difficult to observe in practice (I04). These accounts show that accuracy alone is not the binding constraint; coverage, authenticity, and governance are. Practitioners therefore treat authenticity and provenance as operational, stage-specific quality dimensions, with different acceptance thresholds for training and validation data (I01).

\subsection*{T5: Lawful Governance and Data Rights as a Training-Data Gate}
\textit{The shift: In foundation-model work, lawfulness became an acceptance criterion for data entering the training pipeline.}

Legal compliance, licensing, privacy, and provenance have long shaped how data can be stored, shared, and processed~\cite{khatri2010,simmhan2005}. In foundation-model development, however, lawfulness becomes part of data quality itself. A dataset may be technically useful but still unacceptable for training if its copyright, license, opt-out, or personal-data status is unclear. One participant linked quality directly to data acceptance: quality required “lawful content” with permissive licenses or explicit contracts because “we only include in our LLM data that has checked all these boxes” (I15). The challenge is operationalizing this at scale; the same participant described “the impossibility of knowing what is legal data and what’s illegal data” as a blocker that can paralyze the system (I15).

A second participant made unlearning concrete (I11). Under GDPR a person can demand deletion, but deleting the record does not remove its influence on a trained model. They illustrated this with a linear regression: removing one person's point still leaves a line that interpolates and predicts that person's information, so a court could ask why the system still predicts deleted data, which is why they argued for "provable unlearning processes" (I11). In safety-critical settings, "quality and legal is the same" (I02). These concerns map onto the EU AI Act~\cite{euaiact}, contested copyright in training data~\cite{henderson2023}, and machine unlearning~\cite{bourtoule2021}.

Most evident in foundation-model and agent contexts, lawfulness was treated as a pipeline acceptance criterion, while automotive participants prioritized calibration and representativeness over GDPR.

\subsection*{T6: Representativeness as Behavioral Coverage}

\textit{The shift: Participants treated representativeness as coverage of situations in which the system must behave safely.}

Traditionally, representativeness asks whether data reflects the population of interest~\cite{wang1996}. In safety-critical AI it becomes behavioral coverage: the model must act correctly in situations that are rare, underrepresented, or structurally absent, a concern central to the safety-of-the-intended-functionality standard~\cite{sotif2022}.
A fleet-analytics product owner made this concrete through crash data (I04). The most important crashes are high-severity events, and in exactly those events the data-collection hardware inside the vehicle is destroyed on impact, so the data is never recorded. The team can examine 10,000 crashes and believe it understands crashes, then realize it is "lacking the 10\% most important crashes, which is 95\% of the fatal accidents" (I04). The dataset looks trustworthy and high-quality while systematically missing the cases that matter most. This is not ordinary under-sampling, since the missing data is structurally non-collectable by the system itself.
The risk compounds when gaps become behaviors. An imbalanced regional composition in the training data surfaced as a directional bias in model behavior rather than staying a dataset property (I12). Another participant framed the epistemic limit as "we collect data about the real world, but we don't collect the real world," warning that "edge cases are drowned in the signal of the average" (I14). Representativeness in AI is therefore less about population descriptiveness in the abstract and more about whether the model has been exposed to the conditions under which it must behave safely, extending safety engineering's coverage concern into everyday data-quality practice.

Across all six themes, practitioners repeatedly described conditions that can weaken confidence in data whenever they could no longer explain how data influenced model behavior, who owned it, or whether it remained appropriate for its intended use. Section \ref{sec:discussion} develops our interpretation of this connection between data quality and trust.

\section{Discussion}
\label{sec:discussion}

\subsection{What AI Changes About Data Quality}

This section addresses the three research questions by interpreting the six themes and relating them to prior work. Rather than restating the findings, we draw out their cross-theme implications. Table~\ref{tab:rq-theme-map} summarizes how the themes map to each research question.

For RQ1, practitioners evaluated data quality as fitness for a specific AI claim rather than against a universal checklist. While classic frameworks define quality as fitness for use relative to a consumer and task~\cite{pipino2002,wang1996}, our findings show that AI shifts this judgment to the lifecycle claim the data supports and the evidence available to justify it. Traditional dimensions such as accuracy and completeness~\cite{Batini} therefore remain necessary, but their interpretation changes once data is embedded in model behavior, assessed by models, or constrained by legal rights. Rather than replacing established dimensions, AI extends them into lifecycle assurance.

\input{tables/RQs}

For RQ2, practitioners described data-quality problems that propagate through the AI lifecycle and change form along the way. Defects introduced during collection or preprocessing later reappear as behavioral errors, weak validation confidence, or reduced trust after being mediated by models, synthetic data, or automated quality checks. This interpretation is consistent with data cascades in high-stakes AI~\cite{Sambasivan2021EveryoneWT} and hidden technical debt in production ML~\cite{Hidden_Technical_Debt}, while extending both. Whereas prior work emphasizes neglected data practices or technical dependencies, our participants described propagation even in mature development environments because defects become increasingly difficult to attribute once absorbed into model behavior and AI-generated artifacts.

For RQ3, practitioners described trust as depending on ownership, traceability, and governance across the AI lifecycle. Confidence weakened when practitioners could not identify who was responsible for data quality, how data had been transformed, how it shaped model behavior, or whether its use was permitted. Evidence came from collapsed traceability in end-to-end AI, uncertain authenticity in synthetic or pseudo-labeled data, legal provenance and unlearning concerns, agent context and memory, and missing rare but critical cases. This extends data governance~\cite{khatri2010} and provenance~\cite{simmhan2005} by showing that AI trust requires governing datasets and AI-specific data artifacts.

Taken together, these findings suggest that AI-driven systems shift data quality from a property of datasets to a problem of lifecycle assurance: producing and maintaining evidence that data artifacts are authentic, representative, lawful, traceable, and fit for the AI claims they support.

\subsection{Why the Themes Recur: Five Foundational Explanatory Mechanisms}
\label{sec:foundational_causes}

Looking across the six themes, we observed recurring patterns that we interpreted as reflecting a smaller set of underlying conditions. A second-order interpretation across themes resulted in the five foundational explanatory mechanisms (F1-F5) shown in Fig. ~\ref{fig:causes}.

\input{figures/causal_structure}

\textbf{F1. Data reused beyond its original purpose (the data axis):}
Data collected for one operational purpose is later used for AI it was never specified for, so its quality was never defined against the new use. Participants described data that "was not created for how we use it" (I16) and that carries different quality requirements across consumers (I10). This is the mismatch that dataset documentation aims to prevent~\cite{Datasheets_for_Datasets}, and it drives representativeness gaps (T6) and the turn to synthetic substitutes (T4).

\textbf{F2. AI scale exceeds human inspection (the volume axis):} The volume and modality of training data make exhaustive checking infeasible, so practitioners sample rather than inspect (I05), echoing the ad hoc inspection reported among dataset practitioners~\cite{Understanding_the_Dataset_Practitioners}. The gap pushes toward automated alternatives, driving models as assessors (T2) and pseudo-labeled or synthetic data (T4).

\textbf{F3. Models absorb data into behavior (the model architecture axis):} Data does not remain an external input. It becomes learned behavior, and end-to-end architectures dissolve the boundaries that once supported debugging, the erosion anticipated by work on hidden technical debt~\cite{Hidden_Technical_Debt,zimmerlin2024hiddenbiasesendtoenddriving}. As Theme 1 showed, defects surface as learned behavior rather than inspectable dataset errors (I01, I12). This collapses modular traceability into behavior-level attribution (T1) and turns representativeness gaps into behavioral coverage risks(T6).

\textbf{F4. AI systems create new data artifacts (the new-artifacts axis):}
AI-driven systems introduce data objects produced, judged, or reused by the system itself. These include agent context and memory (I10), model-based data assessment (I01), AI-generated training data (I15), and training data whose lawful use or removal must be demonstrated (I15, I11). Prior work studies these issues separately, such as judge bias, model collapse, and memory governance~\cite{agentmemory2026,shumailov2024,zheng2023}. In our findings, they converge as one assurance problem, explaining why agent context becomes a quality object (T3), models become data-quality assessors (T2), and lawful trainability becomes a quality gate (T5).

\textbf{F5. Organizational ownership of data quality is fragmented (the organizational axis):}
Responsibility for data quality was distributed across sources, transformations, models, and checks, but rarely owned end-to-end. Teams could explain their local stage, while the evidence chain connecting stages often remained no one's responsibility (I03, I04). Data governance defines decision rights and accountabilities for data~\cite{khatri2010}, but our findings suggest that in AI systems these accountabilities can stop at handoffs. This matters because data defects propagate into model behavior, automated checks, and agent actions. Fragmented ownership therefore helps explain traceability loss (T1), circular model-based checking (T2), and unmanaged agent context (T3).

Together, the five mechanisms explain why the themes converge on reduced trust (Fig.~\ref{fig:causes}). Each breaks a different link in the evidence chain: reuse and model absorption obscure lineage (F1, F3), scale limits verification (F2), new AI-generated artifacts escape existing governance (F4), and fragmented ownership leaves the chain without an accountable owner (F5). Because trust in AI data depends on accountable links between sources, transformations, behavior, permitted use, and ownership, each mechanism can weaken it.

\subsection{Implications: Lifecycle Assurance as Evidence Infrastructure}
\label{sec:5.3}

\begin{figure}[t]
    \centering
    \includegraphics[width=1\textwidth]{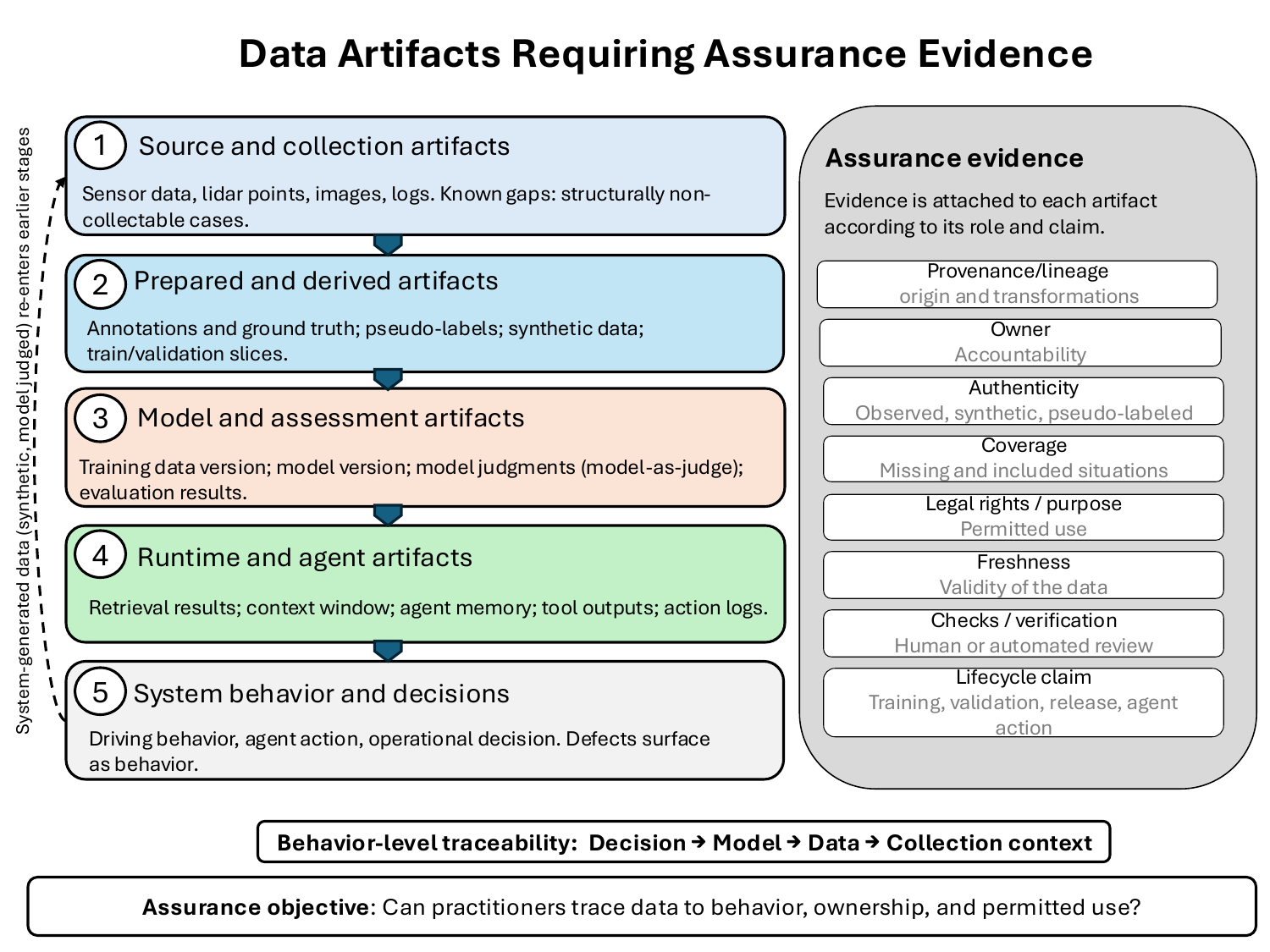}    \caption{Data artifacts and assurance evidence in AI-driven systems: a conceptual model synthesized from the interview findings. Each artifact should be linked to evidence matched to the lifecycle claim it supports, and system-generated artifacts re-enter earlier stages.}
    \label{fig:data-artifacts}
\end{figure}

Across the findings, practitioners' trust weakened whenever they could not connect data to its resulting behavior, ownership, and permitted use. This suggests that data quality can no longer be understood as a property of stored datasets alone. As data becomes embedded in model behavior, assessed by models, and reused through agent context, we synthesize these findings as lifecycle assurance: a conceptual framing in which each data artifact carries the evidence needed to justify the AI claim it supports. Figure~\ref{fig:data-artifacts} summarizes this synthesis by organizing the artifacts described by participants into five lifecycle layers and the evidence associated with each. Because AI-generated artifacts such as synthetic data and model judgments re-enter earlier stages, assurance must also account for data produced by the system itself. The runtime layer extends the emerging agentic contexts described by participants.

Our findings suggest three implications. First, data management should move from generic quality dimensions to \textbf{claim-specific assurance}. Fitness for use ties quality to a consumer and a task~\cite{wang1996}, and our findings narrow it to the claim. A data artifact is not high quality in general but fit for a particular use, such as training, validation, lawful pre-training, or agent action. The question becomes what claim the data supports and what evidence justifies that use.

Second, our findings suggest \textbf{authenticity and provenance} should be treated as first-class metadata. Whether data is observed, synthetic, pseudo-labeled, model-judged, or human-verified determines where it can be trusted and legitimately used. Dataset documentation already treats source datasets this way~\cite{Datasheets_for_Datasets}. We extend the obligation to every artifact in Figure~\ref{fig:data-artifacts}. Provenance audits show such evidence cannot be assumed at scale~\cite{dataprovenance_Attribution_AI}, and regulation increasingly requires it~\cite{euaiact}. Recording origin, verification level, lawful use, freshness, and coverage exposes uncertainty rather than hiding it.

Third, \textbf{accountability should follow the data-model-decision chain}, the traceability path Figure~\ref{fig:data-artifacts} draws from decision back to collection context. Data governance assigns decision rights for quality and access~\cite{khatri2010}, but those rights stop at stage boundaries, which practitioners experienced as fragmented ownership (F5). Organizations may benefit from owning not only individual datasets  but also the evidence chain linking them to model behavior and system decisions.  
Together, these implications frame AI data management as evidence infrastructure: the capability to produce, maintain, and connect evidence about data artifacts across the lifecycle. Its value is not removing uncertainty but managing it, so that data is trusted only within the claims it can support.

\section{Threats to Validity}

This qualitative study seeks analytical rather than statistical generalization. The sample is weighted toward automotive, which may limit transferability to other domains, although this focus provided access to mature data practices in safety-critical settings. Recruitment through professional networks and participant referrals may also have favored practitioners with similar backgrounds. The first author conducted the coding. Consistent with reflexive thematic analysis, we treated coding as interpretive and supported reflexivity through analytical notes, repeated comparison with the full dataset, and discussion among the authors rather than calculating inter-coder agreement. WhisperX transcripts could contain errors, but each transcript was manually checked against the original audio. Finally, lifecycle assurance is an interpretive framing derived from this study and requires further evaluation across organizations and domains.

\section{Conclusion}

This study examined how 16 practitioners across nine organizations define, assess, and manage data quality in AI-driven software-intensive systems. We identified six themes showing that data quality extends beyond dataset correctness to include behavior-level traceability, model-mediated assessment, agent context, authenticity, lawfulness, and behavioral coverage. We further synthesized these themes into five foundational explanatory mechanisms that help account for weakened trust. Our central contribution is a practitioner-grounded framing of data quality as lifecycle assurance, where data quality depends on producing and maintaining evidence that data artifacts are fit for the AI claims they support. Rather than eliminating uncertainty, lifecycle assurance makes uncertainty explicit, reviewable, and accountable across the data-model-decision chain. Our findings suggest that, in AI-driven systems, managing data quality increasingly requires managing the evidence that justifies how data is used, transformed, and trusted throughout the lifecycle.

\textbf{Future work }should evaluate the lifecycle-assurance framing in industrial case studies, investigate how artifact-level assurance evidence can be operationalized in development pipelines, and examine whether it improves data trust and AI outcomes across different application domains.

\textbf{Declaration of AI-assisted technologies in the writing process:} The first author used QuillBot for grammatical correction. All suggestions were reviewed and revised by the first author, who takes full responsibility for the final content.

\begin{credits}
\subsubsection{\ackname}
We would like to thank the Software Center companies for partly funding this research.

\subsubsection{\discintname}
The authors have no competing interests to declare that are relevant to the content of this article.
\end{credits}

\bibliographystyle{splncs04}
\bibliography{references}

\end{document}

%% file: tables/participants.tex
 \begin{table}[t]
\centering
\footnotesize
\setlength{\tabcolsep}{4pt}
\caption{Participants by domain area (16 participants across nine organizations)}
\label{tab:participants}
\begin{tabularx}{\textwidth}{@{}Xl@{}}
\toprule
\textbf{Domain area} & \textbf{Participants IDs} \\
\midrule
Automotive and safety-critical AI & I01 - I07, I12, I13 \\
Telecom and network systems & I08 - I10 \\
Cross-industry AI, data services, and foundation models & I11, I15, I16 \\
Industrial automation & I14 \\
\bottomrule
\end{tabularx}

\vspace{2pt}
{\footnotesize\raggedright Roles across the sample: technical experts and senior technical leaders, deep learning and ML engineers, data and pipeline engineers, data and AI architects, researchers and research managers, a product owner, a quality lead, a data manager, a data expert, and an AI consultant.\par}
\end{table}

%% file: tables/RQs.tex
\begin{table}[t]
\centering
\footnotesize
\setlength{\tabcolsep}{4pt}
\caption{Mapping research questions to themes (T) and foundational mechanisms (F)}
\label{tab:rq-theme-map}
\begin{tabularx}{\textwidth}{@{}lX@{}}
\toprule
\textbf{RQ} & \textbf{Analytical mapping} \\
\midrule
RQ1 & Definition and assessment: from generic quality dimensions to fitness for a specific AI claim (T1, T4, T5, T6) \\
RQ2 & Lifecycle stages: collection and preprocessing (T1, T6); annotation and training (T2, T4); validation (T6); deployment and operation (T2, T3, T5) \\
RQ3 & Trust layers: ownership (F5); traceability (T1); governance of AI data artifacts (T2--T5) \\
\bottomrule
\end{tabularx}
\end{table}

%% file: figures/causal_structure.tex
\begin{figure}[t]
\centering
\resizebox{\columnwidth}{!}{%
\begin{tikzpicture}[
  font=\scriptsize,
  cause/.style={rectangle, rounded corners, draw=black, fill=gray!15,
                text width=2.6cm, align=center, minimum height=0.75cm, inner sep=2pt},
  theme/.style={rectangle, rounded corners, draw=black, fill=blue!10,
                text width=2.8cm, align=center, minimum height=0.75cm, inner sep=2pt},
  trust/.style={rectangle, rounded corners, draw=black, fill=red!12,
                text width=2.9cm, align=center, minimum height=0.8cm, inner sep=3pt},
  primary/.style={-{Stealth[length=1.4mm]}, draw=black!70, thin},
  cross/.style={-{Stealth[length=1.4mm]}, draw=black!45, densely dashed, thin},
]

\node[cause] (f1) {F1. Data reused beyond original purpose};
\node[cause, below=0.18cm of f1] (f2) {F2. AI scale exceeds human inspection};
\node[cause, below=0.18cm of f2] (f3) {F3. Models absorb data into behavior};
\node[cause, below=0.18cm of f3] (f4) {F4. AI systems create new data artifacts};
\node[cause, below=0.18cm of f4] (f5) {F5. Organizational ownership fragmented};

\node[theme, right=2.1cm of f1] (t6) {T6. Representativeness as Behavioral Coverage};
\node[theme, below=0.15cm of t6] (t4) {T4. Authenticity of Generated Data};
\node[theme, below=0.15cm of t4] (t2) {T2. Models as Data-Quality Judges};
\node[theme, below=0.15cm of t2] (t1) {T1. Behavior-Level Traceability};
\node[theme, below=0.15cm of t1] (t3) {T3. Agent Context and Memory as a New Data Object};
\node[theme, below=0.15cm of t3] (t5) {T5. Lawful Governance and Data Rights as a Training-Data Gate};

\node[trust, right=2.0cm of t2] (trust) {Reduced trust in AI data and outputs};

\draw[primary] (f1) -- (t6);
\draw[primary] (f1) -- (t4);
\draw[primary] (f2) -- (t2);
\draw[primary] (f2) -- (t4);
\draw[primary] (f3) -- (t1);
\draw[primary] (f3) -- (t6);
\draw[primary] (f4) -- (t3);
\draw[primary] (f4) -- (t2);
\draw[primary] (f5) -- (t1);
\draw[primary] (f5) -- (t2);
\draw[primary] (f5) -- (t3);
\draw[primary] (f4) -- (t5);

\draw[primary] (t6) -- (trust);
\draw[primary] (t4) -- (trust);
\draw[primary] (t2) -- (trust);
\draw[primary] (t1) -- (trust);
\draw[primary] (t3) -- (trust);
\draw[primary] (t5) -- (trust);

\end{tikzpicture}%
}
\caption{Five foundational explanatory mechanisms connect the six AI data quality themes to reduced trust in AI data and outputs. Arrows denote interpretive analytical links identified in the interview data, not causal claims.}
\label{fig:causes}
\end{figure}